\documentstyle[aps,twocolumn,epsf]{revtex}

\begin{document}
\draft
\title{Approximate Ginzburg-Landau solution for the regular
flux-line lattice. \\
Circular cell method}
\author{W.~V.~Pogosov$^{a}$, K.~I.~Kugel$^{b}$,  A.~L.~Rakhmanov$^{b}$,
        and E.~H.~Brandt$^{c}$}
\address{$^a$ Moscow Institute of Physics and Technology, 141700\\
  Dolgoprudny, Moscow region, Russia\\
  $^b$ Institute for Theoretical and Applied Electrodynamics,\\
  Russian Academy of Sciences, 127412, Moscow, Russia\\
  $^c$ Max-Planck-Institut f\"ur Metallforschung,\\
  D-70506, Stuttgart, Germany}
\date{\today}
\maketitle

\begin{abstract}
A variational model is proposed to describe the magnetic properties
of type-II superconductors in the entire field range between
$H_{c1}$ and $H_{c2}$ for any values of the Ginzburg-Landau
parameter $\kappa>1/\sqrt{2}$. The hexagonal unit cell of the
triangular flux-line lattice is replaced by a circle of the same
area, and the periodic solutions to the Ginzburg-Landau equations
within this cell are approximated by rotationally symmetric solutions.
The Ginzburg-Landau equations are solved by a trial function for the
order parameter. The calculated spatial distributions of the order
parameter and the magnetic field are compared with the
corresponding distributions obtained by numerical solution of the
Ginzburg-Landau equations. The comparison reveals good agreement with
an accuracy of a few percent for all $\kappa$ values
exceeding $\kappa \approx 1$.
The model can be extended to anisotropic superconductors when the
vortices are directed along one of the principal axes. The reversible
magnetization curve is calculated and an analytical formula for the
magnetization is proposed. At low fields, the theory reduces to the
London approach at $\kappa \gg 1$, provided that the exact value of
$H_{c1}$ is used.
At high fields, our model reproduces the main features of the
well-known Abrikosov theory. The magnetic field dependences of
the reversible magnetization found numerically and by our variational
method practically coincide. The model also refines the limits of
some approximations which have been widely used. The calculated
magnetization curves are in a good agreement with experimental data
on high-T$_c$ superconductors.
\end{abstract}


\tighten

\bigskip

\section{Introduction}

The solution to the Ginzburg-Landau (GL) equations found by
Abrikosov \cite {Abrikosov} was used widely \cite{Brandt1,Brandt2}
to study the properties of type-II superconductors at low and high
applied magnetic fields $H$, i.e.~close to the lower and upper critical
fields $H_{c1}$ and $H_{c2}$, respectively. At $H\sim H_{c1}$ the
intervortex spacing is much larger than the vortex core size if
the GL parameter is large, $\kappa \gg 1$. Therefore in the London
model, which is commonly used at low fields \cite{de Gennes}, the
order parameter in the superconductor is assumed to be constant. In
this case, the flux-line lattice (FLL) can be treated as a set of
independent vortices: the magnetic flux density is a linear
superposition of the contributions of individual vortices, and each
contribution coincides with the field of an isolated vortex. The
energy of the system is the sum of the self-energies of the vortices
and their pairwise interaction \cite{Saint James}. At high fields,
the London model looses its applicability because the fraction of
the total volume of superconductor occupied by the vortex cores is
no longer small \cite{Saint James}. Different approximations were
proposed to include the vortex cores and to extend the applicability
limits of the London model \cite{de Oliveira}.

The problem of solving the GL equations for the ideally periodic FLL
can be simplified considerably by replacing the hexagonal unit cell
of the vortex lattice by a circle of equal area
(Wigner-Seitz approximation). In this approach both the order
parameter and the magnetic flux density within the cell have axial
symmetry. The presence of other vortices is taken into account by the
boundary condition: the supercurrent density equals zero at the cell
boundary. This method was used in Ref. \cite{Ihle1}, where an explicit
expression for the magnetization in low fields was found, and the
results are in good agreement with the predictions of the London
model \cite{Saint James,Fetter}. In Refs.
\cite{Ihle2,Kramer,Rammer1,Rammer2,Larkin}, the GL equations and the
equations of the microscopic theory of superconductivity
were solved numerically in the framework
of circular cell approaches and it was shown that this approximation
not only yields good results at low induction but also at
$H \sim H_{c2}$. A similar approach was used
in Ref. \cite{Koshelev}, where the system of GL equations was reduced
to a single equation which can be solved numerically.
The Wigner-Seitz approximation can be extended even to the case of
exotic pairing symmetries, when the order parameter has two complex
components \cite{Barash}.

A {\it numerical} method to find the periodic solutions to the GL
equations was developed in Refs. \cite{Brandt3,Brandt4}. This
exact method accounts for the actual symmetry of the vortex
lattice. It allows calculating the spatial distributions of the
magnetic field and the order parameter within the unit cell, the
elastic shear modulus of the FLL, and the magnetization for any
FLL symmetry and any induction $B$ and GL parameter $\kappa $
with any desired accuracy. However, until recently there was no
adequate approach allowing one to find the magnetization
{\it  analytically} in the entire field range
$H_{c1} <H <H_{c2}$ and to obtain explicit formulas which may
be used to analyze experimental data.

In Ref. \cite{Clem1} Clem proposed a model to solve the GL
equations using a trial function for the order parameter (or
GL function) $|\psi|$:
\begin{equation}  
 |\psi| =f(r) =\frac{r}{\sqrt{r^{2}+\xi _{v}^{2}}}, \label{equil}
\end{equation}
where $\xi _{v}$ is a variational parameter. This model yields an
approximate explicit expression for the local magnetic field of an
isolated vortex:

\begin{equation}  
h(r)=\frac{1}{\kappa \xi _{v}K_{1}(\xi _{v})}K_{0}\left( \sqrt{r^{2}+\xi
_{v}^{2}}\right) ,
\end{equation}
where $K_{n}$ are modified Bessel functions. Here and below the
following dimensionless variables are used: distance $r$, magnetic
flux density $h$, and order parameter $\allowbreak f$ are measured
in units of $\lambda $, $H_{c}\sqrt{2}$, $\sqrt{-\alpha /\beta }$,
respectively, where $\lambda $ is the London penetration depth,
$H_{c}$\ is the thermodynamic critical field, and $\alpha $ and
$\beta $ are the GL coefficients. In this notation, we have
$H_{c2} =\kappa $, $\Phi_{0}=2\pi /\kappa $, where $\Phi_{0}$ is
the magnetic flux quantum. For $\kappa \gg 1$ the minimization of
the free energy gives $\xi _{v}\approx \sqrt{2}/\kappa $ and

\begin{equation}  
 H_{c1}=\frac{1}{2\kappa }(\ln \kappa +\varepsilon ),
\end{equation}
where $\varepsilon $ $=0.52$. The exact value $\varepsilon =0.50$ was
calculated in Ref. \cite{Hu} (see also Ref. \cite{Shapoval}) from the
numerical solution to the GL equations for an isolated vortex.
Note that the lower critical field cannot be found self-consistently
in the framework of the London model, because in this approach the
magnetic flux density diverges on the vortex axis. Therefore, $H_{c1}$
should be regarded as a free parameter in the London expression for the
magnetization \cite{Fetter}. However, the approximation for $H_{c1}$
found by cutting off the field of a vortex at a distance equal to
the coherence length $\xi $, is often used in the London approach
as well \cite{Mitra}. Note that recently Clem's trial function (1)
was applied to the study of the vortex core structure in superconductors
with mixed $(d+s)$ two-component order parameter \cite{Mel'nikov}.

Hao et al. \cite{Hao1} (see also Ref. \cite{Clem2}) extended the
model \cite{Clem1} to larger magnetic fields up to $H_{c2}$ through
the linear superposition of the field profiles of individual vortices.
In this model, the trial function (1) is multiplied by a second
variational parameter $f_{\infty }$  to account for the suppression
of the order parameter due to the overlapping vortex cores. This
model enabled the authors \cite{Hao1} to calculate the magnetization
of type-II superconductors in the full range $H_{c1} < H <H_{c2}$.
Their analytical formula is in a good agreement with the well-known
Abrikosov high-field result. For the case of low fields, Hao and
Clem argued \cite{Hao2} that the London model is quantitatively
incorrect (it does not give the correct asymptotics at $H\to H_{c1}$)
since the contribution of the vortex cores to the total
free energy could not be taken into account in this approach.
The Clem-Hao model was further extended to include
anisotropy \cite{Hao1,Hao3}. This approximation is now widely used
for the analysis of the experimental data on magnetization of
type II superconductors \cite{experiments,exper1,exper2}.

However, it has been recently shown in Ref. \cite{RKP} that the
Clem-Hao model has some drawbacks. It was argued that the procedure
of obtaining the local magnetic flux density by a linear
superposition of contributions of individual vortices in the
form used in Ref. \cite{Hao1} is valid only at low fields. The
application of this approach to the entire field range
$H_{c1}<H<H_{c2}$ leads to an appreciable disagreement
between the Clem-Hao model and Abrikosov's high-field result
\cite{RKP}. In the original papers of Hao et al.
\cite{Hao1,Hao2,Hao3}, this disagreement was made
up by the use of a non-selfconsistent field dependence of the
variational parameters. Then, in calculating the magnetic free
energy in Ref. \cite{Hao1} the lattice sum was approximated by
an integral. As it was shown in Ref. \cite{RKP}, this procedure
leads to a noticeable error in the magnetization at low fields.
This error and the use of an inaccurate value of $H_{c1}$ have
led the authors \cite{Hao1,Hao2} to the conclusion about the
quantitative incorrectness of the London approach at low fields.
Note that in Ref. \cite{Koshelev}
it was argued too that the Clem-Hao
model overestimates the effect of suppression of the order parameter
in the vortex cores at small fields.

In this paper, we propose a variational model for the description of
the regular flux-line lattice in a more consistent fashion as
compared to Ref. \cite{Hao1}. Our variational procedure is based
on Clem's trial function (1). However, in contrast to the Clem-Hao
model, we do not use the superposition of vortex
fields. Instead, we apply the circular cell method and calculate the
magnetic flux density directly from the second GL equation. The
model enables us to find analytical expressions for the local
magnetic field and the order parameter. The results of our variational
procedure are compared with the results of the numerical solution of
the GL equations. This comparison reveals that the analytical formulas
for the spatial distribution of the order parameter and the magnetic
field agree with the numerical results to an accuracy of a few percent
in a wide range of $\kappa \gtrsim 1$ and $B$. By introducing the
effective-mass tensor the theory is extended to include anisotropy
when the vortices are directed along one of the principal axes of
the crystal. The results for the local order parameter and the magnetic
field are then used to calculate the magnetization. The resulting
expression for the magnetization is in agreement with the London model
in small fields at $\kappa \gg 1$, and with the Abrikosov
approximation at $H\sim H_{c2}$. The field dependences of the
magnetization found by the variational and numerical approaches
practically coincide. At the same time, the difference between the
numerical result and the Clem-Hao approach is considerable.
We also found the field dependence of the magnetization in the
Wigner-Seitz approximation close to $H_{c2}$, where the GL equations
can be linearized \cite{Saint James}. The calculated
magnetization curves are compared with the available
experimental data on some high-T$_c$ superconductors.

\section{Theoretical formalism}

In the Wigner-Seitz approximation both the order
parameter and the magnetic flux density within the cell have axial
symmetry. In this case the order parameter $|\psi|$ can be
presented as $ f(r)\exp (-i\varphi )$, with radius vector $r$ and
phase angle $\varphi $. The free energy density
of a superconductor can be written as the sum of two contributions:
$F=F_{em}+F_{core}$. $F_{em}$ is related to the energies of magnetic
field and supercurrent, and $F_{core}$ to the suppression of $|\psi|$
in the vortex core. It is easy to show that in the framework of
the GL theory in the Wigner-Seitz approach $F_{em}$ and $F_{core}$
are given by:

\begin{equation}   
F_{em}=\frac{2\pi }{\kappa A_{cell}}\int_{0}^{R}
\left[ f^{2}\left( a-\frac{1%
}{\kappa r}\right) ^{2}+h^{2}\right] rdr,
\end{equation}

\begin{equation}    
F_{core}=\frac{2\pi }{\kappa A_{cell}}\int_{0}^{R}\left[ \frac{1}{2}%
(1-f^{2})^{2}+\frac{1}{\kappa ^{2}}\left( \frac{df}{dr}\right)
 ^{2}\right]
rdr,
\end{equation}
where $a$ is the dimensionless vector potential,
$R$ and $A_{cell}=\pi R^{2}$ are the cell radius and area, related to
the magnetic induction $B$ by  $A_{cell} = 2\pi /B\kappa $.

The two GL equations can be written as:

\begin{equation}    
-\frac{1}{\kappa ^{2}r}\frac{d}{dr}\left( r\frac{df}{dr}\right)
+f^{3}-f+f\left( a-\frac{1}{\kappa r}\right) ^{2}=0,
\end{equation}

\begin{equation}    
\frac{dh}{dr}=f^{2}\left( a-\frac{1}{\kappa r}\right) .
\end{equation}
The magnetic field and the vector potential are related by

\begin{equation}    
h=\frac{1}{r}\frac{d(ra)}{dr}.
\end{equation}
These equations must be supplemented by the boundary conditions for
the magnetic field and the order parameter:

\begin{equation}    
h(R)=h_{e},
\end{equation}

\begin{equation}    
f(0)=0,\text{ \ \ \ \ \ \ }f'(R)=0,
\end{equation}

\begin{equation}    
rf^{-2}(r)\frac{dh}{dr}=-1/\kappa \text{ \ \ \ \ at\ \ }\ r\to 0.
\end{equation}
Condition (11) follows from Eqs.~(7) and (8) \cite{Saint James}.
The system of Eqs.~(6)-(11) is much simpler than the similar equations
for the hexagonal unit cell. However, even this system can be solved
only numerically \cite{Ihle2}. Nevertheless, in high and small fields
some results can be obtained analytically. At small fields, the
approximate solution to the GL equations in the Wigner-Seitz approach
was found in Ref. \cite{Ihle1}. Here, the spatial variation of the
order parameter at $\kappa $ $\gg 1$ can be neglected when calculating
the magnetic flux density. In this case, $h$ can be found analytically
from the second GL equation (6) and the boundary conditions (9) and
(11). This yields the magnetization \cite{Ihle1}:

\begin{equation}   
 -4\pi M(B)=H_{c1}+\frac{1}{2\kappa }\left[ \frac{K_{1}(R)}{I_{1}(R)}
 +\frac{1}{2I_{1}^{2}(R)}\right] -B.
\end{equation}
When $H\gg $ $H_{c1}$, the radius of the cell is $R\ll 1$, and
Eq.~(12) can be expanded in powers of $R$, yielding:

\begin{equation}  
 -4\pi M(B)=H_{c1}-\frac{1}{4\kappa }\left[ \ln (2\kappa
  (H-H_{c1})+\sigma \right] ,
\end{equation}
with $\sigma =1.3456$. A similar relationship was obtained in
the London limit in Ref. \cite{Fetter} for the regular FLL, with
$\sigma =1.3431$ \cite{Fetter} for the triangular lattice.

At high fields, the magnetization is given
by \cite{Saint James}:

\begin{equation}   
M=\frac{H-H_{c2}}{4\pi \beta _{A}(2\kappa ^{2}-1)},
\end{equation}
where $\beta _{A}$ depends only on the symmetry of the
FLL \cite{Saint James}:
\begin{equation}   
 \beta_{A}=\frac{\int f^{4}d^{2}{\bf r}}{\left[
 \int f^{2}d^{2}{\bf r}\right] ^{2}} \,.
\end{equation}
Here the integrals are taken over the area of the unit cell.
Let us find the value of $\beta _{A}$\ for the circular cell.
Near $H_{c2}$, the magnetic field undergoes only slight spatial
variation. The vector potential in this case is $a \approx Br/2$.
The order parameter is small at $H\sim H_{c2}$, so the first GL
equation (6) can be linearized. The resulting equation has the
analytical solution \cite{Abramowitz}:

\begin{equation}   
 f(r)=s\exp \left( -\frac{\kappa rB}{2}\right) \Phi \left(
 \frac{B-\kappa }{2} , 2 , \frac{\kappa rB}{2}\right) ,
\end{equation}
where $\Phi $ is the Kummer function. Factor $s$ depends on
the nonlinear term in Eq.~(6) but it does not affect the value of
$\beta _{A}$. Using Eqs.~(15) and (16), we find $\beta_A=1.1576$
for the circular cell. This value is close to $\beta_A=1.1596$
calculated in Ref. \cite{Kleiner} for the triangular lattice.

Thus, in both limits of low and high fields, the magnetization in
the Wigner-Seitz approximation is in good agreement with that for
the regular triangular FLL, which has the lowest energy. In the next
section, we propose a variational model to solve the GL equations
in the whole field range between $H_{c1}$ and $H_{c2}$ at any
$\kappa >1/\sqrt{2}$ with good accuracy.

\section{Variational procedure}

Using the trial function (1) multiplied by a variational prefactor,
$f(r) =f_{\infty } r / \sqrt{r^{2}+\xi _{v}^{2}}$, allows us to solve
the second GL equation (7) analytically within the Wigner-Seitz cell:

\begin{equation}    
 h(r)=uI_{0}(f_{\infty }\sqrt{r^{2}+\xi _{v}^{2}})+vK_{0}(f_{\infty }
 \sqrt{r^{2}+\xi _{v}^{2}}) ,
\end{equation}
where $u$ and $v$ can be found from the boundary conditions
(9) and (11):
\begin{equation}    
u=\frac{f_{\infty }}{\kappa \xi _{v}}\frac{K_{1}(f_{\infty }\rho )}{%
K_{1}(f_{\infty }\xi _{v})I_{1}(f_{\infty }\rho )-I_{1}(f_{\infty }\xi
_{v})K_{1}(f_{\infty }\rho )},
\end{equation}
\begin{equation}    
v=\frac{f_{\infty }}{\kappa \xi _{v}}\frac{I_{1}(f_{\infty }\rho )}{%
K_{1}(f_{\infty }\xi _{v})I_{1}(f_{\infty }\rho )-I_{1}(f_{\infty }\xi
_{v})K_{1}(f_{\infty }\rho )},
\end{equation}
and we introduced the notation $\rho =\sqrt{R^{2}+\xi _{v}^{2}}$.
Note that, rigorously speaking, the Clem trial function (1) does not
meet the condition (10) since its derivative can not be equal to zero
at the cell boundary. However,  $df/dr$ is small at $r=R$ and the
comparison between the results of variational and numerical methods
demonstrates good accuracy of the approach.

The values of variational parameters should be found by minimization
of the total free energy density $F=F_{em}+F_{core}$. Using
Eqs.~(7) and (4), it is possible to obtain the following expression
for the magnetic energy density: $F_{em}=Bh(0)$ \cite{Hao1}. Taking
into account Eqs. (17)-(19), we get:

\begin{eqnarray}  
F_{em} &=&\frac{Bf_{\infty }}{\kappa \xi _{v}}\times  \nonumber \\
&&\times \frac{K_{0}(f_{\infty }\xi _{v})I_{1}(f_{\infty }\rho
)+I_{0}(f_{\infty }\xi _{v})K_{1}(f_{\infty }\rho )}{K_{1}(f_{\infty }
\xi_{v})I_{1}(f_{\infty }\rho )-I_{1}(f_{\infty }\xi _{v})
  K_{1}(f_{\infty }\rho )}.
\end{eqnarray}
The energy related to the spatial variation of the order parameter
in the vortex core is found from Eq.~(5) by a straightforward
integration \cite{Hao1,Hao3}:\bigskip
\begin{eqnarray}   
F_{core} &=&\frac{1}{2}(1-f_{\infty }^{2})^{2}+  \nonumber \\
&&+\frac{1}{2}B\kappa \xi _{v}^{2}f_{\infty }^{2}(1-f_{\infty }^{2})
\ln \left[ 1+\frac{2}{B\kappa \xi _{v}^{2}}\right] +  \nonumber \\
&&+\frac{f_{\infty }^{4}}{2}-\frac{f_{\infty }^{4}}{2+B\kappa
\xi _{v}^{2}} +\frac{Bf_{\infty }^{2}\left( 1+B\kappa \xi _{v}^{2}
\right) }{\kappa \left( 2+B\kappa \xi _{v}^{2}\right) ^{2}}.
\end{eqnarray}
\newline

The field dependence of the variational parameters is calculated
numerically by minimization of the total free energy
$F(B,\kappa ,\xi _{v},f_{\infty })$ with respect to $\xi _{v}$ and
$f_{\infty }$. This dependence can be approximated by the following
analytical expressions with an accuracy of about 0.5\%:

\begin{eqnarray}   
\xi _{v}(B,\kappa ) &=&\xi _{v0}\times  \nonumber \\
&&\times \left( 1-4.3\left( 1.01-\frac{B}{1.05\kappa }\right) ^{6.3}
\left(\frac{B}{\kappa }\right) \right) ^{1/2}\times  \nonumber \\
&&\times \left( 1-0.56\left( \frac{B}{\kappa }\right) ^{0.9}
\right) ^{1/2},
\end{eqnarray}

\begin{eqnarray}   
f_{\infty }(B,\kappa ) &=&\left( 1-\frac{B^{2}}{2.8\kappa ^{2}}\right)
\times \left( 1-\left( \frac{B}{t\kappa }\right) ^{4}\right) ^{1/2}
\nonumber \\
&&\times \left( 1+\frac{1.7B}{\kappa }\left( 1-\frac{1.4B}{\kappa }
\right)^{2}\right) ^{1/2},
\end{eqnarray}
where $t=0.985$, $\xi_{v0}$ is the value of $\xi_{v}$ at $B=0$. The
latter can be calculated from the condition $dF/d\xi_{v}=0$ at $B=0$:
\begin{equation}   
\kappa \xi _{v0}=\sqrt{2}\left[ 1-\frac{K_{0}^{2}(\xi _{v0})}
{K_{1}^{2}(\xi_{v0})}\right] .
\end{equation}
When $\kappa \gg 1$, Eq.~(24) has the solution
$\xi _{v0}=\sqrt{2} /\kappa $ ($\sqrt{2}\xi$ in dimensional variables).
Eqs.~(1), (17)-(19), (23), and (24) give the distributions of the
order parameter and the magnetic field within the Wigner-Seitz cell.

\begin{figure}[tbp]
\epsfxsize= .95\hsize  \vskip 1.0\baselineskip
\centerline{ \epsffile{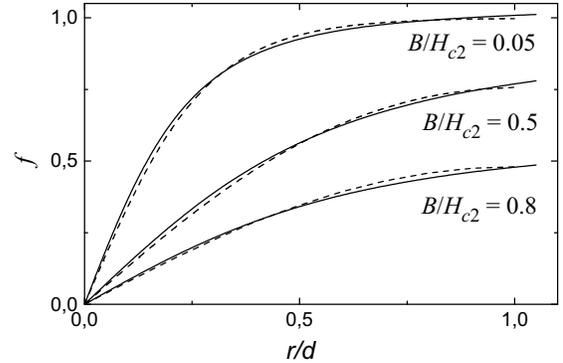}}
\caption{The spatial distribution of the dimensionless order parameter
in the unit cell of the flux-line lattice at different magnetic
inductions $B=0.1H_{c2}$, $B=0.5H_{c2}$, $B=0.8H_{c2}$ for
$\protect\kappa =10$.
The solid lines correspond to the variational calculations in the
Wigner-Seitz approximation. The dashed lines correspond to the
numerical solution of the Ginzburg-Landau equations for the triangular
lattice (nearest neighbor vortices are in the plane of the graph).
The distance is measured in units of the intervortex spacing $d$ in
the triangular lattice.}
\label{fig1}
\end{figure}

The upper critical field is determined as the field at which the
order parameter in the superconductor becomes equal to zero. As can
be seen from Eq.~(23), one has $f=0$ at $H=0.985\kappa $. Thus, the
difference between the exact $H_{c2}=\kappa $ and its calculated
value is about 1.5\%. This result is quite natural, since variational
procedures in general give only approximate solutions to the GL
equations. Similarly, the Clem value of $H_{c1}$ slightly differs
from the numerically calculated one \cite{Hu,Shapoval}.

Now we compare the obtained results for the order parameter and
the magnetic field with the similar distributions computed for the
triangular lattice by the numerical method proposed in
Ref. \cite{Brandt4}. The dependence of the
order parameter on the distance from the cell center is shown in
Fig.~1 for different values of the magnetic induction at $\kappa =10$.
The spatial distribution of the order parameter in the triangular
lattice along the nearest neighbor direction is also shown
in Fig.~1. The results of our variational calculations are close
to the numerical ones at any values of the magnetic induction. The
difference does not exceed several percent. Such an accuracy of
our approach remains in a wide range of $\kappa \gtrsim 1$. The magnetic
flux density as a function of the distance from the vortex axis is
shown in Fig.~2 at $\kappa =10$, $B=0.5H_{c2}$. There is good
agreement between the variational and numerical results for $h$.
In Fig.~3, the spatial average of the order parameter squared,
 $\omega = \langle |\psi|^2 \rangle $,
is plotted as a function of the magnetic induction at
$\kappa =100$. The comparison with the numerical result shows
that the deviation of variationally-calculated $\omega $ from the
exact dependence does not exceed one percent in small and
intermediate fields. Near $H_{c2}$ this deviation increases due to
a small difference between the calculated $H_{c2}$ and the exact value.

\begin{figure}[tbp]
\epsfxsize= .95\hsize  \vskip 1.0\baselineskip
\centerline{ \epsffile{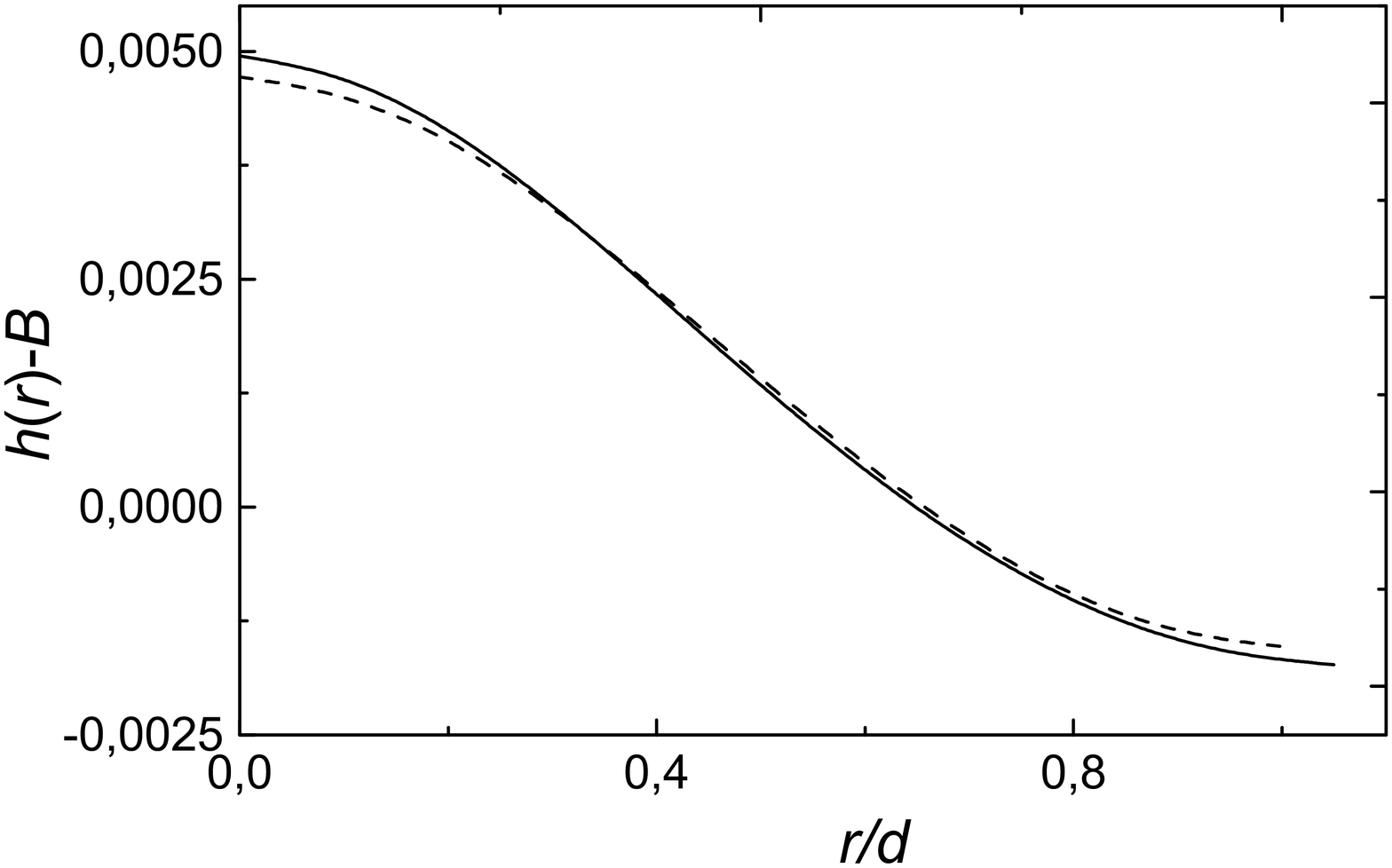}}
\caption{The spatial distribution of the dimensionless magnetic field
in the unit cell of flux-line lattice at the magnetic induction
$B=0.5H_{c2}$ for $\protect\kappa =10$. The solid line corresponds to
the variational calculations in the Wigner-Seitz approximation.
The dashed line corresponds to the numerical solution.}
\label{fig2}
\end{figure}

Thus, the results of our variational approach agree well with the
exact numerical solution to the GL equations. In the the next
section we shall apply this approach to the calculation of the
magnetization curve.

\begin{figure}[tbp]
\epsfxsize= .95\hsize  \vskip 1.0\baselineskip
\centerline{ \epsffile{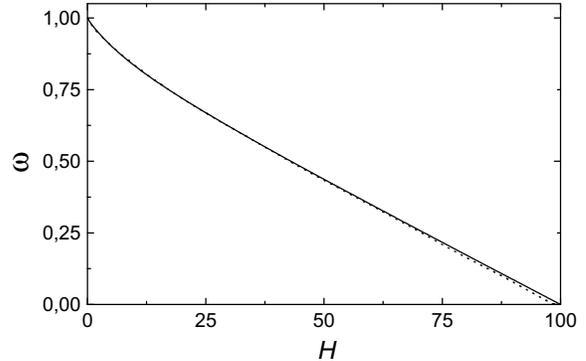}}
\caption{Averaged order parameter squared $\protect\omega$ vs $B$ in
dimensionless units for $\protect\kappa=100$. The solid line
corresponds to the variational calculations and the dot line
to the numerical solution.}
\label{fig3}
\end{figure}

The anisotropy in the GL functional can be taken into account by
introducing the phenomenological effective-mass tensor
$m_{j}$ ($j=1,2,3$), where $m_{j}$ are the effective masses in the
direction of the principal axes $x_{j}$. It was shown in
Ref. \cite{Kogan} that the Ginzburg-Landau equations can be
transformed to isotropic form by a simple transformation if $\kappa $ is
replaced by $\kappa ^{\ast }=\kappa p_{1}^{-1/2}$, where
$p_{1}=m_{1}/\sqrt{m_{1}m_{2}m_{3}}$,  (the vortices are directed
along the $x_{1}$ axis). Thus, in this case,
the Wigner-Seitz approximation can be used. The Wigner-Seitz
cell has an elliptic shape and can be transformed to the circular
cell by a scaling transformation: the distance along the $x_{j}$ axis
should be normalized by $m_{1}m_{2}m_{3}/\sqrt{m_{1}m_{j}}$. The
magnetization of the anisotropic superconductor can be found from the
magnetization of the isotropic one by replacing the GL parameter
$\kappa $ by $\kappa^*$.

\section{Magnetization}

\bigskip The magnetization is defined by the well-known relationship:
\begin{equation}   
 -4\pi M=H-B.
\end{equation}
It can be calculated by two equivalent methods if the exact solution
to the GL equations is known. First, the magnetic field $H$ may be
calculated by minimization of the Gibbs free energy $G=F-2BH$:

\begin{equation}   
 H=\frac{1}{2}\frac{\partial F}{\partial B}
\end{equation}
This derivative was calculated, e.g.~in Refs. \cite{Fetter,Hao1}.
The second approach uses the virial theorem for the
flux-line lattice, which was proven in Ref. \cite{Doria}, namely the
applied magnetic field $H$ can be found from the local magnetic field
$h$ and the order parameter $f$ as

\begin{equation}   
H=\frac{1}{2BA_{cell}}\int \left( f^{2}-f^{4}+2h^{2}\right) d^{2}
{\bf r},
\end{equation}
where the integral is taken over the area of the unit cell. Both
methods are equivalent if the exact solutions $f$ and $h$
are used \cite{Doria}.

The variational model gives the spatial distributions of the order
parameter and the magnetic field within the cell, which are close to
the exact results. Let us find the magnetization by means of both
methods. According to Eq.~(26) the magnetization $M$ is:

\begin{eqnarray}
-4\pi M &=&-B+\frac{f_{\infty }}{\kappa \xi _{v}}\times  \nonumber \\
&&\times \frac{K_{0}(f_{\infty }\xi _{v})I_{1}(f_{\infty }\rho
)+I_{0}(f_{\infty }\xi _{v})K_{1}(f_{\infty }\rho )}{K_{1}(f_{\infty }\xi
_{v})I_{1}(f_{\infty }\rho )-I_{1}(f_{\infty }\xi _{v})K_{1}(f_{\infty }\rho
)}+  \nonumber \\
&&+\frac{1}{2B\kappa ^{2}\xi _{v}^{2}}\allowbreak \{K_{1}(f_{\infty }\xi
_{v})I_{1}(f_{\infty }\rho )-  \nonumber \\
&&-I_{1}(f_{\infty }\xi _{v})K_{1}(f_{\infty }\rho )\}^{-2}+\frac{f_{\infty
}^{2}\left( 2+3B\kappa \xi _{v}^{2}\right) }{2\kappa \left( 2+B\kappa \xi
_{v}^{2}\right) ^{3}}+  \nonumber \\
&&+\frac{\kappa ^{2}f_{\infty }^{2}\xi _{v}^{2}}{2}\{\frac{1-f_{\infty }^{2}%
}{2}\ln \left[ \frac{2}{B\kappa \xi _{v}^{2}}+1\right] +  \nonumber \\
&&+\frac{f_{\infty }^{2}\left( 2+3B\kappa \xi _{v}^{2}\right) }{2\kappa
\left( 2+B\kappa \xi _{v}^{2}\right) ^{3}}-\frac{f_{\infty }^{2}\left(
2+3B\kappa \xi _{v}^{2}\right) }{2\kappa \left( 2+B\kappa \xi
_{v}^{2}\right) ^{3}}\}.
\end{eqnarray}
Thus, in the former case the dependence of the magnetization on the
magnetic induction $B$\ is given by Eqs.~(28) and (22)-(24). The
dependence of $H$ on $B$ is given by Eq.~(25). Thus, we find the
implicit function $M(H)$.

 Within the second approach, the integral (27) can be calculated only
numerically when $f$ and $h$ are defined by Eqs.~(1) and (17). Our
calculations show that not only the values of $H$ found by both
methods coincide, but also the values of $M$, which is usually much
smaller than $H$, are practically indistinguishable. The difference
between them is much less than one percent at any induction and
$\kappa >1/\sqrt{2}$. Note that the result for $M$ in the second
approach is very sensitive to the perturbations of $f(r)$ and
$h(r)$. For example, if one puts  $f_{\infty }=1$ near $H_{c1}$ and
minimizes the free energy only with respect to $\xi _{v}$ this does
not change $f(r)$ and $h(r)$ considerably. However, this procedure
would lead to an appreciable change of $M(H)$ when using
Eq.~(27), while according to Eq.~(26) the magnetization practically
remains the same. Below, we shall use Eq.~(28) for the magnetization.

At low fields the variational parameters (22) and (23) may be
considered as constants independent of $B$ when $\kappa \gg 1$. In
this case, Eq.~(28) can be expressed as a power series in terms of
$\xi_{v}$.  As a result, it is possible to obtain Eq.~(12) with
$H_{c1}$ given by Eq.~(3) at $\varepsilon $ $=0.52$. Thus, in small
fields the model reduces to the London approximation provided that
the variationally-calculated value of $H_{c1}$ is used, which is
practically indistinguishable from the exact $H_{c1}$. Actually,
the use of the exact $H_{c1}$ in small fields is equivalent to
taking into account the effect of vortex cores. The field dependence
of the magnetization (28) is shown in Fig.~4 for $\kappa =100$.
The magnetization curves corresponding to the London and Abrikosov
approximations are also plotted. At low fields, the magnetization
practically coincides with the results of the London approach, the
difference between them does not exceed 0.5\%. At $H\sim H_{c2}$,
the behavior of the magnetization is in good agreement with the
Abrikosov high-field result (14). Although near $H_{c2}$ both curves
are close to each other, the error of our approximation is not
so small as in low fields because of the slight deviation of the
calculated $H_{c2}$ from the exact value. For example, at
$H=0.8H_{c2}$, $\kappa =100$  the error of our variational procedure
is about 5\%. In order to improve the accuracy near $H_{c2}$, one
may put the constant $t$ in Eq.~(23) to be equal to 1. As a result,
the difference between the variational and the Abrikosov $M(H)$
curves decreases in the vicinity of $H_{c2}$, whereas at low and
intermediate fields the magnetization does not change.

In the inset of Fig.~4, we compare the calculated dependence
$-4\pi M(H)$ with that found in Ref. \cite{RKP}, where the Clem-Hao
approach of superposition of vortex fields \cite{Hao1} was used.
It is clearly seen that this approach is valid only at small
fields, and its use leads to an appreciable error in the
magnetization even in the intermediate field range. An additional
error in $M$ arises in small fields due to the approximate
replacement of the lattice sums by integrals \cite{Hao1} in
calculating the magnetic free energy; for more details see
Ref. \cite{RKP}.

\begin{figure}[tbp]
\epsfxsize= .95\hsize  \vskip 1.0\baselineskip
\centerline{ \epsffile{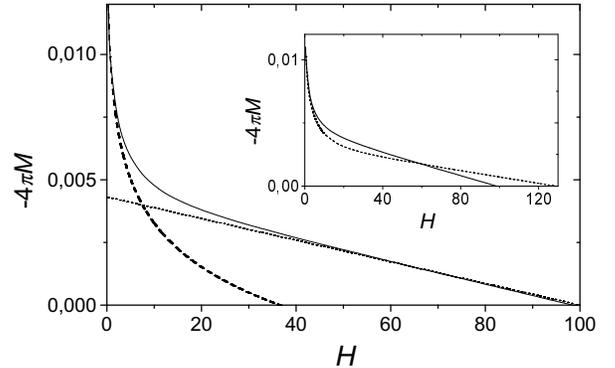}}
\caption{Calculated $-4\protect\pi M(H)$ using the variational method
for $\protect\kappa =100$ (solid line). Also shown are the London
dependence $-4\protect\pi M(H)$ (13) with the exact value of
$H_{c1}$ (3) (dashed line) and the Abrikosov high-field result (16)
(dot line). The inset refers to the magnetization found within
the framework of the variational model (solid line) and the result
of calculations \protect\cite{RKP} for the magnetization in the
Clem-Hao approximation \protect\cite{Hao1} (dot line). The
difference between these curves arises due to the superposition of
vortex fields used in the Clem-Hao model \protect\cite{Hao1}.
Dimensionless variables are used. }
\label{fig4}
\end{figure}

Now we compare the magnetization found in the framework of the
variational and numerical methods. At low and high fields the exact
dependence $M(H)$ coincides with the results of the London (at $\kappa \gg 1$)
 and the Abrikosov approximations, respectively. As we found above, the
results of the variational approach are in agreement with these
approximations. In the intermediate field range, where the London
and the Abrikosov approaches are not applicable, the difference
between the values of the magnetization calculated by numerical and
variational methods is not bigger than 1\% in a wide range of $\kappa \gg 1$
values. Thus, our results for the magnetization appear to be a good
approximation to the exact numerical solution of the GL equations at
$\kappa \gg 1$.

Next we discuss the case of small $\kappa$ values. In Fig.~5,
the field dependences of the magnetization are plotted for several
small $\kappa$ values. The solid and dotted lines correspond to the
variational and numerical calculations, respectively. The agreement
between these results is good
at $\kappa \gtrsim 1$. At smaller $\kappa$ values the variational and
the exact numerical results differ near the lower critical field.
In this case, the intervortex distance is of the order of the
coherence length almost in the entire field range, and the variational
approach based on an appropriate trial function for the order
parameter in the circular Wigner-Seitz cell may lead to some
deviation from the exact solution.

In the inset
of Fig.~5 we compare the magnetization calculated by means of our
variational procedure and by the Clem-Hao model at $\kappa =1$.
It is seen that also for small (even as for large $\kappa$,
see above) the calculation method proposed in Ref. \cite{Hao1}
leads to an inaccurate magnetization.

\begin{figure}[tbp]
\epsfxsize= .95\hsize  \vskip 1.0\baselineskip
\centerline{ \epsffile{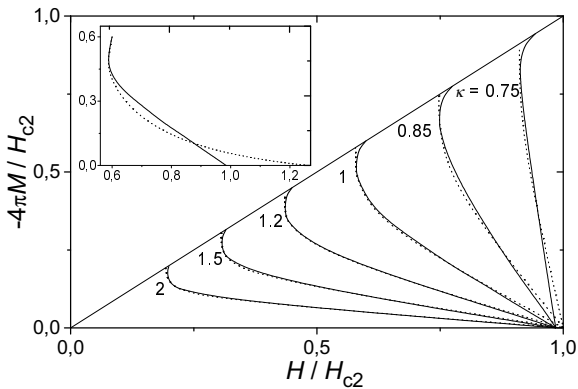}}
\caption{Magnetization curves calculated using the variational approach
(solid lines) and exact numerical method (dotted lines) at various
small $\kappa$ values. The inset compares the magnetization obtained
by our variational model (solid line) with the Clem-Hao
approximation \protect\cite{Hao1} (dashed line) at $\protect\kappa =1$.
 Dimensionless variables are used.}
\label{fig5}
\end{figure}

Our formulas for the magnetization may be used for the analysis
of experimental data. In Fig.~6 the calculated magnetization curves
are compared with the measured magnetization of
YBa$_{2}$Cu$_{4}$O$_{8}$ polycrystals \cite{exper1} and
Nd$_{1.85}$Ce$_{0.15}$CuO$_{4-\delta }$
single crystals \cite{exper2}.  In these papers, the magnetization
curves at different temperatures were analyzed and reduced to
the dimensionless form based on the Clem-Hao formulas with
non-selfconsistent field dependences of the variational
parameters \cite{Hao1}. The resulting magnetization curve is close
to the Abrikosov high-field result and to our variational dependence
in the intermediate field range. The $\kappa$ values obtained in Refs. \cite{exper1,exper2} were
 $\kappa =70$ for YBa$_{2}$Cu$_{4}$O$_{8}$ and $\kappa =80$
for Nd$_{1.85}$Ce$_{0.15}$CuO$_{4-\delta }$.

\begin{figure}[tbp]
\epsfxsize= .95\hsize  \vskip 1.0\baselineskip
\centerline{ \epsffile{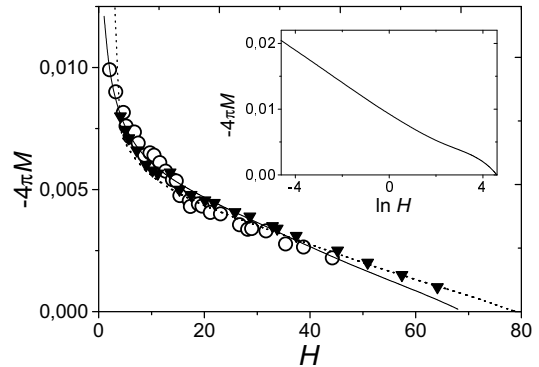}}
\caption{The field dependence of magnetization in dimensionless
units. The solid and dotted lines show the theoretical variational
dependences at $\protect\kappa =70$ and $\protect\kappa =80$. The circles and
triangles give the experimental data for
YBa$_2$Cu$_4$O$_8$ \protect\cite{exper1} and
Nd$_{1.85}$Ce$_{0.15}$CuO$_{4-\delta}$ \protect\cite{exper2}, respectively.
The inset shows the magnetization calculated by our variational model
and plotted versus the logarithm of the applied field $H$ at
 $\protect\kappa =100$.}
\label{fig6}
\end{figure}

In Fig.~6 we compare these experimental curves with our
variational result. The circles and triangles in Fig.~6 correspond to
YBa$_{2}$Cu$_{4}$O$_{8}$ and Nd$_{1.85}$Ce$_{0.15}$CuO$_{4-\delta }$,
respectively. The solid and dotted lines correspond to the
theoretical curves at $\kappa =70$ and $\kappa =80$, respectively.
It is clearly seen that good agreement exists between theory and
experiment. In the inset of Fig.~6 the theoretical (variational)
magnetization is plotted as a function of $\ln H$ at $\kappa =100$.
This dependence is nearly linear in a wide range of intermediate
fields. A similar behavior of the magnetization was observed in
numerous experiments; see, for example, Ref. \cite{Mitra}.

\section{Conclusions\qquad \qquad}

We proposed an approximate method to solve the Ginzburg-Landau
equations for the regular flux-line lattice at any values of the
magnetic induction and the Ginzburg-Landau parameter
$\kappa >1/\sqrt{2}$. The Wigner-Seitz approximation is used, and
the hexagonal unit cell of the vortex lattice is replaced by a circle
with the same area. Our model is based on Clem's trial function for the
order parameter. The use of this function allows us to find the
magnetic flux density self-consistently from the second
Ginzburg-Landau equation. The comparison between the variational
results and the results of exact numerical solution of the
Ginzburg-Landau equations reveals good accuracy of our approach:
the difference between the spatial distributions of the order
parameter and the magnetic field does not exceed several percent. Such
accuracy remains in a wide range of values of the magnetic induction
and $\kappa \gg 1$. The method is applied to the calculation of the field
dependence of the reversible magnetization. An analytical expression
for the magnetization is proposed. At low fields, the obtained
dependence agrees with the predictions of London theory at
$\kappa \gtrsim 1$. At high fields, it is in good agreement with the
 Abrikosov result. It is shown that the values of the magnetization
calculated within the framework of our variational model and of the
numerical method of solution of the Ginzburg-Landau equations are
practically indistinguishable, especially in small and intermediate
magnetic fields at $\kappa \gtrsim 1$. Our model yields the limits of
the Clem-Hao model for the magnetization. The presented
analytical formulas for the magnetization may be used to analyze
experimental data. As an illustration we compared the experimental
and calculated magnetization curves for different high-T$_c$
superconductors  (YBa$_{2}$Cu$_{4}$O$_{8}$ and
Nd$_{1.85}$Ce$_{0.15}$CuO$_{4-\delta }$) and found good agreement
between theory and experiment.

\section*{ACKNOWLEDGMENTS}

The authors acknowledge useful discussions with L.\ G.\ Mamsurova,
K.\ S.\ Pigalskiy, and N.\ G.\ Trusevich. This work is supported by
the Russian Foundation for Basic Research (RFBR), grants
\#00-02-18032 and  \#00-15-96570, by the joint INTAS-RFBR program,
grant \#IR-97-1394, and by the Russian State Program
 'Fundamental Problems in Condensed Matter Physics'.


\begin{references}
\bibitem{Abrikosov}  A. A. Abrikosov, Zh. Eksp. Teor. Fiz. {\bf 32},
1442 (1957) [Sov. Phys. JETP {\bf 5}, 1174 (1957)].

\bibitem{Brandt1}  E. H. Brandt and U. Essman, Phys. stat. solidi
 (b) {\bf 144}, 13 (1987).

\bibitem{Brandt2}  E. H. Brandt, Rep. Prog. Phys. {\bf 58}, 1465 (1995).

\bibitem{de Gennes}  P. G. de Gennes, {\em Superconductivity
of Metals and Alloys} (Benjamin, New York, 1966).

\bibitem{Saint James}  D. Saint-James, G. Sarma, and E. Thomas,
{\em Type-II Superconductors} (Oxford, Univ. Press, New York, 1969).

\bibitem{de Oliveira}  I. G. de Oliveira and A. M. Thompson,
Phys. Rev. B {\bf 57}, 7477 (1998).

\bibitem{Ihle1}  D. Ihle, Phys. stat. solidi (b) {\bf 47}, 423 (1971).

\bibitem{Fetter}  A. L. Fetter, Phys. Rev. {\bf 147}, 153 (1966).

\bibitem{Ihle2}  D. Ihle, Phys. stat. solidi (b) {\bf 47}, 429 (1971).

\bibitem{Kramer}  R. J. Watts-Tobin, L. Kramer, and W. Pesch, J. Low
Temp. Phys. {\bf 17}, 71 (1974).

\bibitem{Rammer1}  J. Rammer, W. Pesch, and L. Kramer,
Z. Phys. B {\bf 68}, 49 (1987).

\bibitem{Rammer2}  J. Rammer, J. Low Temp. Phys. {\bf 71}, 323 (1988).

\bibitem{Larkin}  A. I. Larkin and Yu. N. Ovchinnikov, Phys. Rev. B
{\bf 51}, 5965 (1995).

\bibitem{Barash} A. S. Mel'nikov, Zh. Eksp. Teor. Fiz. {\bf 101}, 1979 (1992)
 [JETP {\bf 74} (6), (1992)].

\bibitem{Brandt3}  E. H. Brandt, Phys. stat. solidi (b)
{\bf 51}, 345 (1972).


\bibitem{Brandt4}  E. H. Brandt, Phys. Rev. Lett. {\bf 78}, 2208 (1997).

\bibitem{Clem1}  J. R. Clem, J. Low Temp. Phys. {\bf 18}, 427 (1975).

\bibitem{Hu}  C.-R. Hu, Phys. Rev. B {\bf 6}, 1756 (1972).

\bibitem{Shapoval}  E. A. Shapoval, JETP Letters {\bf 69}, 577 (1999).

\bibitem{Mitra}  V. G. Kogan, M. M. Fang, and S. Mitra,
Phys. Rev. B {\bf 38}, 11958 (1988).

\bibitem{Koshelev}  A. E. Koshelev, Phys. Rev. B {\bf 50}, 506 (1994).

\bibitem{Mel'nikov}  A. S. Mel'nikov, I. M. Nefedov, D. A. Ryzhov,
I. A. Shereshefskii, and P. P. Vysheslavtsev, Phys. Rev. B
{\bf 62} (2000).

\bibitem{Hao1}  Z. Hao, J. R. Clem, M. W. McElfresh, L. Civale, A. P.
Malozemoff, and F. Holtzberg, Phys. Rev. B {\bf 43}, 2844 (1991).

\bibitem{Clem2}  J. R. Clem, in {\em Superconducting Electronics},
edited by H. Weinstock, and M. Nisenoff (Springer -Verlag,
Berlin, 1989), p.1.

\bibitem{Hao2}  Z. Hao and J. R. Clem, Phys. Rev. Lett {\bf 67},
2371 (1991).

\bibitem{Hao3}  Z. Hao and J. R. Clem, Phys. Rev. B {\bf 43},
7622 (1991).

\bibitem{experiments}  J. H. Gohng and D. K. Finnemore,
Phys. Rev. B {\bf 46}, 398 (1992); Q. Li {\em et al.}, Phys. Rev. B
{\bf 46}, 3195 (1992); D. N. Zheng {\em et al.},
Phys. Rev. B {\bf 48}, 6519 (1993); V. C. Kim {\em et al.}, Phys. Rev. B
{\bf 51}, 11767 (1995); P. Pugnat {\em et al.}, Europhys. Lett.
{\bf 29}, 425 (1995); M.-S. Kim {\em et al.}, Phys. Rev. B {\bf 51},
3261 (1995); M. Xu {\em et al.}, Phys. Rev. B {\bf 53}, 15313
(1996); Y. Zhuo {\em et al.}, Phys. Rev. B {\bf 55}, 12719 (1997);
M.-S. Kim {\em et al.}, Phys. Rev. B {\bf 57}, 6121 (1998);
M.Y. Cheon {\em et al.}, Physica C {\bf 302}, 215 (1998);
S.-I. Lee {\em et al.}, Physica C {\bf 341}, 379 (2000);
H.-J. Kim {\em et al.}, Physica C {\bf 341}, 745 (2000);
A. Aburto {\em et al.}, Physica C {\bf 303}, 185 (1998); H.-J. Kim
{\em et al.}, Physica C {\bf 331}, 292 (2000).

\bibitem{exper1}  W. Chen {\em et al.}, Phys. Rev. B {\bf 51}, 6035 (1995).

\bibitem{exper2}  A. Nugroho {\em et al.}, Phys. Rev. B {\bf 60}, 15384
(1999).

\bibitem{RKP}  W. V. Pogosov, A. L. Rakhmanov, and K. I. Kugel,
Zh. Eksp. Teor. Fiz. {\bf 118}, 908 (2000) [JETP {\bf 91}, 588 (2000)].

\bibitem{Abramowitz}  M. Abramowitz and F. A. Stegun, {\em Handbook of
Mathematical Functions} (Dover, New York, 1965).

\bibitem{Kleiner}  W. M. Kleiner, L. M. Roth, and S. H. Autler,
Phys. Rev. A {\bf 133}, 1226 (1964).

\bibitem{Kogan}  V. G. Kogan and J. R. Clem, Phys. Rev. B {\bf 24}, 2497
(1981).

\bibitem{Doria}  M. M. Doria, J. E. Gubernatis, and D. Rainer,
Phys. Rev. B {\bf 39}, 9573 (1989).
\end{references}
\end{document}